\documentclass[prd,twocolumn,showpacs,preprintnumbers,amsmath,amssymb,superscriptaddress,nofootinbib]{revtex4}
\usepackage{amsmath}
\usepackage{amssymb}
\usepackage{amsthm}
\usepackage{overpic,color}
\usepackage{afterpage}
\usepackage{epsfig}
\usepackage{float}
\usepackage{rotating}
\usepackage{lscape}
\usepackage{enumerate}
\usepackage{xspace}
\usepackage{url}
\usepackage{upgreek}
\usepackage{dcolumn}      
\usepackage{bm}           
\usepackage{multirow}
\usepackage{bigstrut}
\usepackage{upgreek}
\usepackage[tight]{units} 
\begin{document}

\title{Snowmass 2013 Computing Frontier: Intensity Frontier}         

\author{B.~Rebel}
\affiliation{Fermi National Accelerator Laboratory, Batavia, IL 60510, USA}
\author{M.~C.~Sanchez}
\affiliation{Iowa State University, Ames, IA, 50010 USA}
\affiliation{Argonne National Lab, Lemont, Illinois 60439, USA}
\author{S.~Wolbers} 
\affiliation{Fermi National Accelerator Laboratory, Batavia, IL 60510, USA}

\date{\today}

\begin{abstract}
The Intensity Frontier (IF) is a primary focus of the U.S.-based particle physics program. It encompasses a large spectrum of physics, including quark flavor physics,  charged lepton processes, neutrinos, baryon number violation,  new light weakly-coupled particles, and nucleons, nuclei and atoms. There are many experiments, a range of scales in data output and throughput, and a wide range in the number of experimenters. The experiments, projects and theory in this area all require demanding computing capabilities and technologies.  The IF experiments have significant computing requirements for simulation,  theory and modeling, beam line and experiment design, triggers and DAQ, online monitoring, event reconstruction and processing, and physics analysis. We have conducted a qualitative survey of the current and near-term future experiments in the IF to understand the computing demands of this area and their expected evolution. This report details the expected computing requirements for the IF in the context of the Snowmass Community Summer Study 2013. 
\end{abstract}

\maketitle

\section{Introduction}
\label{sec:intro}
Computing at the Intensity Frontier (IF) has many significant challenges. The experiments, projects and theory all require demanding computing capabilities and technologies.  Though not as data intensive as the LHC experiments, the IF experiments and IF computing have significant computing requirements in theory and modeling, beam line and experiment design, triggers and DAQ, online monitoring, event reconstruction and processing, and physics analysis.  It is critical for the success of the field that IF computing is modern, capable, has adequate capacity and support, and is able to take advantage of the latest developments in computing hardware and software advances.

This report will detail the computing requirements for the IF.  In \S~\ref{sec:overview}, a short overview of the the IF will be given.  An understanding of the IF program now and in the future is necessary to appropriately discuss the computing associated with IF.  The next section will discuss computing for the IF.  The emphasis here will be on the experiments, specifically on the aspects of computing required for IF experiments including the data handling architecture.  The following section summarizes information from IF experiments, current and planned, collected in a recent survey.  Next will come a discussion of some of the issues involved in computing for IF. This will include beam line design, simulation, demands for detector design, demands on Geant4, with particular emphasis on the importance of keeping up with current computing technology and techniques and aligning with the energy frontier computing whenever possible.  Finally there will be a summary.

\section{Overview of the Intensity Frontier: Recent Growth and Future Prospects}
\label{sec:overview}

The IF encompasses: $i)$ quark flavor physics, $ii)$ charged lepton processes, $iii)$ neutrinos, $iv)$ baryon number violation, $v)$ new light weakly coupled particles, and $vi)$ nucleons, nuclei and atoms~\cite{snowmass-if-document}. The requirements and resources of quark flavor physics, as in Belle II and LHCb, are more similar to those of the energy frontier. The requirements and resources of $iv)$ and $v)$ are more similar to those of the cosmic frontier. We have thus maintained focused on the areas of charged lepton processes, neutrinos, baryon number violation and nucleons, nuclei and atoms. 

The IF has become the central focus of the US-based particle physics program.  The transition to the IF dominated domestic program coincides with the transition at Fermilab from operating Energy Frontier (EF) experiments to operating IF experiments.  Many of the IF experiments are designed to measure rare processes by using very intense beams of particles.  Successful running of these experiments will involve not only the delivery of high intensity beams, but also the ability to efficiently store and analyze the data produced by the experiments. 

Several experiments comprise the Fermilab-based IF, including experiments to measure neutrino cross sections (MiniBooNE, MicroBooNE, MINER$\nu$A), experiments to measure neutrino oscillations over long (MINOS$+$, NO$\nu$A, LBNE) and short baselines (MiniBooNE, MicroBooNE), experiments to measure muon properties ($g-2$, $\mu2e$), other precision experiments (SEAQUEST), as well as future experiments (ORKA, $\nu$STORM).  Each of those experiments represent collaborations between 50 and 400 people.

There is also strong US participation in several international IF experiments, such as Super-Kamiokande (SK), T2K, Daya Bay, SNO/SNO+ as well as US university lead experiments such as IceCube. The impact of the US contribution to the physics results of these experiments is strongly correlated to the availability of computing resources and the efficiency of the computing model adopted. The groups participating in these experiments range in size from 30 to 250 people. In addition there is significant detector and experiment design R\&D. 

The IF has a large number of experiments and a range of scales in data output and throughput as well as number of experimenters. The situation is thus very different than at the EF which has just few experiments each of a very large scale. The number of experiments and range of scales can potentially lead to fragmentation, reinvention of the wheel, lack of access to computing advances and in general more dollars for each type of computing and personnel support needed. Furthermore, while there might be significant overlap of the human resources among experiments, there is little benefit when the tools and other resources used diverge significantly. A broad range of experiments leads to a wide breadth of needs, from support of tens to hundreds of experimenters, from high intensity realtime processing but small data sets to large data sets which in the sum are equivalent to the previous generation of collider experiments. Over the last few years there has been a significant effort by the IF experiments at Fermilab to join forces in using a more homogeneous set of software packages, frameworks and tools to access infrastructure resources. This trend has reduced fragmentation and led to more efficient use of resources. We would like to see this trend expand in the broader IF community adapted to the needs of each collaboration. 

\section{Computing for the Intensity Frontier}
\label{sec:computing}

\subsection{Survey of Current and Future Computing Needs}
\label{sec:survey}

A qualitative survey was conducted of the current and near term future experiments in the IF in order to understand their computing needs and also the foreseen evolution of these needs.  Computing liaisons and representatives for the LBNE, MicroBooNE, MINER$\nu$A, MINOS$+$, Muon $g-2$, NO$\nu$A, SEAQUEST,  Daya Bay, IceCube, SNO+, Super-Kamiokande and T2K collaborations all responded to the survey. This does not cover all experiments in all areas but we consider it a representative survey of the IF  field.

The responses and conclusions to the survey can be summarized in five aspects: $i)$ support for software packages, $ii)$ support for software frameworks, $iii)$  access to dedicated and shared resources, $iv)$ access to data handling and storage, and overall computing model and its evolution.

\subsubsection{Support for Software Packages}
\label{sec:external}

All of the responding experiments listed ROOT~\cite{root} and Geant4~\cite{geant4} as software packages that are critical to their operation and as such must continue to be supported.  Geant4 has traditionally focused on EF experimental support, stronger ties and support with the IF experiments is thus a requirement. As an example of this need, Geant4 was cited in the survey as barely suitable in speed for simulating large scintillation detectors, given a relatively complex geometry and a large number of photons to track. The IF experiments have started to work with some members of the Geant4 collaboration to properly tune Geant4 for the energy regime of those experiments and to determine appropriate parameter settings.  The funding of dedicated people to work on these projects would enable continued improvement of the simulations and better results.  Several respondents indicated a desire for improved efficiency in the Geant4 package. The Geant4 collaboration is testing multi-threading modifications of the code at this time and it will soon become available more widely. 

In addition to these primary packages, many of the US neutrino experiments use the state-of-the-art GENIE~\cite{genie} package for simulating neutrino interactions in their detectors.  The neutrino community will benefit from expanded support of the GENIE efforts as well as broader involvement in developing the modules that comprise the different interaction processes in GENIE.  It is a convenient framework which allows the development of independent models for simulation. An interface from Geant4 to GENIE is also in development and will be of great benefit to the IF neutrino experiments.

Other key packages, such as FLUKA~\cite{fluka} are used to simulate the production of hadrons in beam line simulations as well as nuclear interactions in detectors.  The support of beam line simulation packages should not be neglected. In addition, the community uses or has developed a variety of specialized physics packages used, for example CRY~\cite{cry} for simulating cosmic rays and  NEST~\cite{nest} for determining ionization and light production in noble liquid detectors. 

Theory also plays a vital role in advancing the goals of the IF, including the development of software tools used to determine optimal experimental configurations, performing global fits of available data, and developing phenomenological models.  An example of a tool used by several long baseline neutrino experiments during their planning phases is GLoBES~\cite{Huber:2007ji} which provides expected physics reach given input parameters such as the beam spectrum and expected efficiency of the proposed detector.  Similar efforts are needed to combine the wealth of data produced in the current and next generation of neutrino detectors.  An effort to develop a toolkit for doing that should be supported~\cite{pmns-toolkit}. Muon experiments could also benefit from support of the theory community to develop models and predictions.

There is significant benefit to encouraging collaborative efforts among experiments.  For example, LArSoft~\cite{larsoft} is a common simulation, reconstruction and analysis toolkit for use by experiments developing simulations and reconstructions for liquid argon time projection chambers (LArTPCs). The LArSoft package is managed by Fermilab.  All US experiments using LArTPCs currently use LArSoft.  Similarly, the NuSoft~\cite{nusoft} toolkit is being used by the LArTPC experiments in the US as well as by NOvA.  This project is a joint effort between those experiments. Such efforts make better use of development and maintenance of resources.

A related need amongst all experiments is a reliable and stable event display toolkit.  MINERvA and MicroBooNE are using a web based utility called Arachne~\cite{Tagg:2011wk} that appears to be very promising.  Other event displays in use by IF experiments are based on the ROOT GUI library.  It would be helpful if a common toolkit could be identified and supported.

In addition to these simulation and reconstruction related software packages, there are several other computing tools that are widely used in the field that should be maintained.  These tools provide infrastructure access for code management, data management, grid access, electronic log books and document management. Experimental use of these tools is more varied than with the specialized ones, unless the experiments are based at Fermilab. Examples of these tools include the Electronic Collaboration Logbook (ECL) and Document Database (DocDB) projects developed and maintained at Fermilab.  Similarly, Indico and Redmine~\cite{redmine} are broadly used for arranging meetings and storing documentation.

\subsubsection{Software Frameworks}
\label{sec:frameworks}

A broad range of experiments typically write or use many different frameworks that must be supported. The significant overlap of experimenters across experiments makes this a large overhead on the training, support and expertise development. Efforts for common frameworks can have a significant impact. 

The Fermilab-based IF experiments (g-2, $\mu2$e, NO$\nu$A, ArgoNeuT, LArIAT, MicroBooNE, LBNE) have converged on ART as a framework for job control, I/O operations, and tracking of data provenance. This framework is developed and maintained by the Scientific Computing Division at Fermilab by computing professionals. It has perhaps the largest user base within IF at this time. 

Increased resources for the ART framework could enable some of the needs experiments such as more accessible parallelization of experimentÕs code, for example using standard thread libraries (OpenMP, TBB). In fact, the primary limitation listed by users of ART was the inability to parallelize jobs at the level of individual algorithms.  The ability to do so will become more critical as the numbers of channels in IF experiments continue to increase and the separation of signal from background becomes more difficult due to the rare nature of the processes being examined.  This ability will also allow IF experiments to take advantage of the design of modern computers containing multiple cores.

Experiments outside of Fermilab (or within Fermilab but that decided on a framework before ART was available) use LHC derived frameworks such as Gaudi or homegrown frameworks like MINOS(+), IceTray and RAT. The level of support for development and maintenance of such frameworks varies depending if the experiment is a significant stakeholder and/or significant human resources are available. These experiments would also benefit from more accessible parallellization and professional computing support.

Additionally, Fermilab-based experiments responding to the survey indicated that they use computing professionals at Fermilab either as consultants for software development, as is the case fore experiments using ART, or directly in the development of data acquisition programs, as is the case for NO$\nu$A and MicroBooNE.  Every experiment indicated that if more computing professional effort were made available, they could efficiently make use of that effort to accomplish $i)$ parallelization of code, $ii)$ establishing offsite batch submission farms, $iii)$ establishing best practices for writing software, $iv)$ software development, and $v)$ optimizing use of Geant4. The availability of this expertise is in high demand within the IF community and the already existing expertise at Fermilab could fulfill this need to the wider IF community both inside and outside of Fermilab if this was promoted and funded. 

\subsubsection{Access to dedicated and shared resources}

Typically the hardware demands of IF experiments are modest compared to those of the EF experiments.  However, that does not mean that the needs are insignificant.  For example, each experiment foresees the need of at least 1000 dedicated slots for submitting jobs to batch processing facilities.  Other interesting requirements are:
\begin{itemize}
\item NO$\nu$A projects needing 4.8 million CPU hours per year to produce simulation files alone.
\item LBNE expects to need several PB of storage space each year during operation.
\item Even smaller scale experiments like MINER$\nu$A and MicroBooNE expect to use PB of storage.
\end{itemize}

The current and projected CPU needs for the IF experiments are shown in Table~\ref{table:cpu}.    The CPU needs are given in terms of the number of dedicated grid slots needed for each experiment each year, and the estimated sum for US-based and international-based experiments are also given. These estimates are based on near-term needs and involve some projections which should be taken only as an order of magnitude. It should be noted that the grid usage by IF experiments tends to follow a feast and famine pattern.  The production of simulation, reconstruction of simulation and data, and the analysis of those files follows cycles that are strongly correlated with the major conference cycle.  The peak usage per experiment should be used in determining the needs rather than the steady state usage since the ideal peak time usage can reach ten times the planned steady state usage. To ensure the ability to meet the peak usage needs, each experiment should have a dedicated number of slots that is a large fraction, at least 50\%, of the typical peak usage need as well as access to run opportunistically on a much larger pool of slots. Having that level of resources available ensures timely production of results. 

\begin{table*}
\caption{\label{table:cpu}  Annual dedicated grid slot usage by experiment in the IF.  The total columns call out the totals for the Fermilab-based and foreign-based experiments separately.  The experiments based outside of the US are only shown as the total.} 

\begin{tabular}{|l|c|c|c|c|c|c|c|c|c|} 

\hline \hline

Year  &  NO$\nu$A  & MicroBooNE & LBNE & g-2    & MINER$\nu$A & MINOS$+$ & $\mu2e$ & Total Fermilab & Total International\\ \hline 
2013 & 1000            & 500               & 200     & 500   & 1200                & 1200          & 1800        &   6400               & 1250             \\ \hline
2014 & 1500            & 750               & 400     & 1000 & 1400                & 1200          & 2300        &   8550               & 1500             \\ \hline
2015 & 2000            & 1000             & 800     & 1000 & 1600                & 1200          & 2500        & 10100             & 2000            \\ \hline
2016 & 2500            & 1000             & 1500   & 1000 & 1800                & 1200          & 2500        & 11500             & 2500            \\ \hline
2017 & 2500            & 1000             & 2000   & 1000 & 1800                & 1200          & 2500        & 12000             & 2500            \\ \hline
2018 & 2500            & 1000             & 2500  & 1000 & 1500                & 1000          & 2500        & 12000             & 2500            \\ \hline
2019 & 3000            & 750               & 2500   & 1000 & 1000                & 750            & 3000       &  12000             & 2500           \\ \hline
2020 & 3000            & 500               & 3000   & 1000 & 1000                & 500            & 3000        & 12000             & 3000           \\ \hline
2021 & 3000           & 500               & 3000   & 1000 & 1000                & 500            & 3000        &  12000             & 3000           \\ \hline
2022 & 3000            & 250               & 3500   & 1000 & 1000                & 250            & 3000        & 12000            & 3000          \\ \hline
2023 & 3000            & 100               & 3800   & 1000 & 1000                & 100            & 3000        & 12000             & 3000          \\ \hline
\hline \hline

\end{tabular}

\end{table*}

There is excellent support of the Fermilab based experiments both in terms of storage and CPU.  Issues are mostly in efficient data handling and script optimization. Resources for computing professionals is provided through Fermilab and would be extremely useful if increased. On site grid access is however not sufficient, offline Monte Carlo generation is common among experiments. Professional support is thus required for methods to seamlessly use Fermilab and non-Fermilab resources through job submission protocols. For Fermilab-based experiments, university and other national lab resources are used in the production of Monte Carlo files. A common protocol to access these resources such as OSG is in the foreseeable future.

The IF experiments in which US physicists participate but are not based in the US have significantly less support.  Even though they are recognized by the Open Science Grid, US groups have no dedicated US-based grid computing resources. These experiments tend to rely either on resources in other countries, with low priority, or on university based resources that are shared amongst a broad pool of university users from multiple disciplines.  As an example experiments like T2K run intensively on grid resources in Europe and Canada.  Canadian and UK grid support was cited several times as a model both for grid computing and grid storage. These researchers must have access to dedicated resources that can be shared with other IF experiments in order to be competitive with analysis of data and simulation. It was widely noted that the lack of dedicated US resources has a detrimental impact on the science. 

The IF computing networking requirements are that the data be able to be move easily to the necessary locations, be accessible for data acquisition, reconstruction, simulation and analysis and that there is the ability to take advantage of distributed computing, either as part of the grid or cloud.  The networking must not be a barrier to making effective use of the distributed computing that is available and allow collaborations to reconstruct and analyze the data in a distributed way.  The scale of the networking need is estimable by comparing the scale of the IF computing to the EF computing. As IF moves to larger and more international collaborations the network requirements will be will grow as the experiments collect more data and more people analyze the data and the people are more distributed.

\subsubsection{Access to data handling and storage}
\label{sec:datahandling}

Respondents from experiments based at Fermilab indicated that their primary data copies are stored at Fermilab.  The infrastructure there handles active storage as well as archiving of data.  The SAM system designed and maintained at Fermilab was noted as the preferred data distribution system for these experiments. Heavy I/O for analysis of large numbers of smaller sized events is an issue for systems like BlueArc. Fermilab should continue to receive support from the DOE to ensure proper archiving of data. Other experiments indicated using grid protocols for data storage. 

All respondents indicated the need for data handling systems that seamlessly integrate distribution of files across the network from multiple locations.  This desire enables experiments to make optimal use of national lab and university resources.  The need for such a system is acutely felt by experiments that are not based at Fermilab.  One possible solution to this problem could resemble the tiered computing structure used by the LHC experiments, with all IF experiments making use of that structure.

The current and projected storage needs for the IF experiments are shown in Table~\ref{table:storage}.  The storage needs are presented as the additional storage needed each year. These are estimates based on near term projected needs by the experiments and extrapolation based on the planned life cycles of the various experiments at Fermilab. 

\begin{table*}
\caption{\label{table:storage}  Additional annual permanent storage (in terabytes) needed by each experiment in the IF.  The total columns call out the totals for the Fermilab-based and foreign-based experiments separately.  The experiments based outside of the US are only shown as the total.} 

\begin{tabular}{|l|c|c|c|c|c|c|c|c|c|} 

\hline \hline

Year  &  NO$\nu$A  & MicroBooNE & LBNE & g-2    & MINER$\nu$A & MINOS$+$ & $\mu2e$ & Total Fermilab & Total Foreign\\ \hline 
2013 & 1020            & 120               & 20       & 20     & 580                 & 90               & 30          & 1880                 & 750              \\ \hline
2014 & 3020            & 270               & 40       & 20     & 580                 & 90               & 30          & 5070                 & 750              \\ \hline
2015 & 5020            & 270               & 80       & 50     & 580                 & 90               & 30          & 6120                 & 1000            \\ \hline
2016 & 5020           & 270               & 160     & 350   & 580                 & 90               & 30          & 6500                 & 1000            \\ \hline
2017 & 5020            & 270               & 320     & 350   & 580                 & 90               & 30          & 6660                 & 1300            \\ \hline
2018 & 5020             & 120               & 900    & 350   & 200                 & 40               & 30          & 6660                 & 1300            \\ \hline
2019 & 5020             & 60                 & 1200   & 350   & 100                 & 40               & 40          & 6810                 & 1300            \\ \hline
2020 & 5020             & 60                 & 1800   & 350   & 100                 & 20               & 40          & 7390                 & 1500            \\ \hline
2021 & 5020             & 60                 & 2500   & 450   & 50                   & 20               & 300        & 7380                 & 1500            \\ \hline
2022 & 5020             & 20                 & 2500   & 450   & 50                   & 10               & 300        & 7430                 & 1700            \\ \hline
2023 & 5020            & 10                 & 3000   & 450   & 50                   & 10               & 300        & 7820                 & 2000            \\ \hline
\hline \hline

\end{tabular}

\end{table*}

\subsubsection{Overall Computing Model}
\label{sec:compmodel}

The respondents indicated a high degree of commonality when describing their experiment's computing model despite large differences in the type of data being analyzed, the scale of that processing, or the specific workflows followed.  The model is summarized as a traditional event driven analysis and Monte Carlo simulation using centralized data stores that are distributed to independent analysis jobs running in parallel on grid computing clusters.  In the current model of provisioning, there is a remarkable overlap in the infrastructure used by experiments. For large computing facilities such a Fermilab, it would be useful to design a set of scalable solutions corresponding to each of these patterns, with associated toolkits that would allow access and monitoring. Provisioning an experiment or changing a computing model would then correspond to adjusting the scales in the appropriate processing units.

The consensus is that computing should be made transparent to the user, such that non-experts can perform any reasonable portion of the data handling and simulation.  Moreover, all experiments would like to see computing become more distributed across sites, but only in very large units where it can be efficiently maintained.  Users without a home lab or large institution require equal access to dedicated resources. 

The evolution of the computing model follows several lines including taking advantage of new computing paradigms, like clouds, different cache schemes, GPU and multicore processing.  Perhaps even more importantly, we need to continuously make improvements in reducing the barrier of entry for new users, make the systems easier to use, and add facilities that help prevent the users from making mistakes.

In regards to computing technology, there is a concern that as the number of cores in CPUs increases, RAM capacity and memory bandwidth will not keep pace, causing the single-threaded batch processing model to be progressively less efficient on future systems unless special care is taken to design clusters with this use case in mind. 

There is currently no significant use of multi-threading, since the main bottlenecks are Geant4 (single-threaded) and file I/O. Geant4's multithreading addition might have a very significant impact across the field. There is also a possibility of parallelization at the level of the ART framework. Greater availability of multi-core/GPU hardware in grid nodes would provide motivation for upgrading code to use it. For example currently we can only run GPU-accelerated code on local, custom-built systems. 

\subsection{Designing Future Intensity Frontier Experiments}
\label{sec:design}

The design of IF experiments requires massive amounts of simulation both for the detectors and the beam lines.  The beam line design work for LBNE, $\mu2$e and g-2 has used approximately 20\% of the grid slots used by all IF efforts at Fermilab.  The simulation of new detectors also requires intensive grid use in order to test the different configurations for optimizing the design.  Naturally, intensive use of the grid translates into intensive use of storage, which drives the storage usage even before experiments start taking data. 

There are also challenges in the realm of data acquisition, integrated low-power chips with multiple CPU cores are the new standard. The next generation filter processing must be performed in this multi-core environment with smaller local cache memory for each computing core. Potential sensitivity gains could be achieved by the high level processing of complex events, for example factor of three for ORKA over the previous generation of kaon experiments. Work must be done towards streaming acquisition architectures relying on high levels of filter processing, as high as a factor of 1000. This will require close integration with offline computing and a robust computing framework to reap the benefits~\cite{instrumentation-IF-report}. 

Next generation experiments will also have to address the open data policies now being established by the funding agencies. A  clear avenue for sharing multi-TB or PB data samples, an acceptable format, and guidance on how we release these to the public must be established. Additional resources to support these data sets once they are released and curate these samples over the long term will be needed. Efforts by a world-wide group DPHEP and within the EF to address this issue will be of great benefit to the IF experiments. 

\subsection{Computing Technology for the Intensity Frontier and Commonalities with the other Frontiers}
\label{sec:tech}

Naturally the IF experiments share the need of continued development of common HEP tools like Geant4 and ROOT with the other frontiers.  In addition, the IF experiments will heavily benefit from the development of a data handling system that is easily distributed and has transparent access for the user, as mentioned in \S~\ref{sec:datahandling}.  The example of the tiered computing used by the LHC experiments could be a good basis for developing the model for the IF.  It is expected that any solution developed would provide access to the data as well as methods for submitting jobs to the grid. Advances in adapting key software tools to exploit multi-threading and GPU environments will also be beneficial across frontiers. 

We highlight three efforts whose benefits might cut across frontiers, the Chroma package from IF, ATLAS simulations on HPCs from EF and self-assembling data from CF. 

Chroma~\cite{chroma} is an open source optical photon Monte Carlo which simulates standard physics processes such as diffuse and specular reflections, refraction, absorption, Rayleigh scattering, and scintillation. Photons are propagated in parallel on many-core modern GPUs. It can propagate 2.5 million photons/sec in a large detector with 29,000 PMTs. This is roughly 200 times faster than the same simulation with Geant4. Integration of a similar system into Geant4 would produce large gains for IF and possibly other experiments as GPUs become available in more standard HTC environments. 

An effort is ongoing to run ATLAS simulations on ANL's Intrepid~\cite{lecompte}. Using these machines in ATLAS will require a front-end which accepts jobs to OSG, starts the job, does the initialization and db access, and then accepts the output and finalizes the job. The strategy of usage is backfilling. While there is general agreement that there might not be enough idle time in HPCs, improvements in the code and enabling access is the first step towards using more modern architectures for HEP simulations. 

Finally, the problems of packaging, transporting, and processing large volumes of data in real time are of critical importance throughout the frontiers. For the Cosmic Frontier, CTA must gather approximately 30 GB/s of data from 100 telescopes distributed over a km$^2$ area and process it in real time so that observation strategies can be modified in response to transient phenomena. A multidisciplinary group at the CF proposes to design a fault-tolerant real-time association of information across their large-scale experiment containing distributed sensors by creating a self-assembling data paradigm~\cite{weinstein}. Future applications to more transparent and efficient data distribution can be envisioned with this approach.

\section{Summary}
\label{sec:comp-summary}

The computing needs of the IF experiments should be viewed collectively.  When combined, these experiments require the resources and support similar to a single EF experiment.  The support of these experiments directly impacts the quality of results and the efficiency with which those results can be obtained.  There is significant support already for IF experiments that are based at Fermilab and the required support there is expected to increase as the current generation of experiments under construction begin to take data.  The support of IF experiments that are not based at Fermilab but still have significant US collaboration, such as T2K, needs to be improved.  Specifically, there should be an investment in infrastructure and professional support to serve these experiments.  

The Computing Frontier should also strive for transparent access to data and hardware resources for the IF.  Users must have a simple interface with which to request data sets that then determines the stored location of those data and returns the data quickly to the user.  Similarly, there should be a standardized grid submission tool that determines the optimal location for running jobs without the user having to specify.  

The IF benefits significantly from the ability to share common frameworks and tools, such as ART, GENIE, NuSoft and LArSoft.  The support of these efforts must be continued and increased as new experiments come on line and more users are added to current experiments.  Similarly, the common tools used across all frontiers, such as ROOT and Geant4, must be supported and continuously improved. Computing professionals are in demand as support for key software frameworks, software packages, scripting access to grid resources and data handling. Fermilab is a natural center for IF support in these areas given the existing expertise and large number of IF experiments already on site. 

There are efforts (and problems) that are shared across frontiers, significant investments in ROOT and Geant4 optimizations, HPC for HEP, transparent OSG access and open data solutions would have a high payoff. 

\bibliography{ifcomputing_summary}

\end{document}